\pdfoutput=1
\documentclass[aps,prd,amsmath,floats,floatfix,twocolumn,
superscriptaddress,nofootinbib,showpacs,longbibliography]{revtex4-1}

\usepackage{packages}

\begin{document}

\title{Emergent structure in the binary black hole mass distribution and implications for population-based cosmology}

\author{Vasco Gennari\,\orcidlink{0000-0002-0190-9262}}
\email{vasco.gennari@l2it.in2p3.fr}
\affiliation{Université de Toulouse, CNRS/IN2P3, L2IT, Toulouse, France}

\author{Tom Bertheas\,\orcidlink{0009-0005-4118-4170}}
\affiliation{Université de Toulouse, CNRS/IN2P3, L2IT, Toulouse, France}
\affiliation{Laboratoire de Physique de l'ENS, Université Paris Cité, Ecole Normale Supérieure, Université PSL, Sorbonne Université, CNRS, 75005 Paris}

\author{Nicola Tamanini\,\orcidlink{0000-0001-8760-5421}}
\affiliation{Université de Toulouse, CNRS/IN2P3, L2IT, Toulouse, France}

\hypersetup{pdfauthor={Gennari et al.}}

\date{\today}

\begin{abstract}

Gravitational waves provide a powerful probe of both the astrophysical processes driving black hole mergers and the dynamics of the Universe, but these measurements rely on accurately inferring the unknown underlying population. We perform an agnostic reconstruction of the primary mass distribution using B-splines, characterising the emergence of structure with increasing model complexity. Using the latest gravitational-wave transient catalog, GWTC-4.0, we identify multiple mass features and find evidence suggesting a logarithmic hierarchy in the population. We show that this structure directly impacts measurements of the Hubble constant, primarily through features at the population boundaries. Finally, we introduce an approach that isolates a subpopulation of low-mass events to mitigate modelling systematics, providing a promising path toward robust population-based cosmology with future datasets.

\end{abstract}

\maketitle

\section{Introduction}
\label{sec:introduction}

Ten years after the first direct detection of \acp{GW}~\cite{LIGOScientific:2016aoc}, the large number of observed \ac{BH} mergers has made \ac{GW} physics a flourishing field driven by observations~\cite{Callister:2024cdx, Mastrogiovanni:2024mqc, Berti:2025hly}. Among the numerous science cases, a robust comprehension of the population that governs our observations is fundamental not only to understanding the underlying astrophysical processes that make up \acp{BBH}~\cite{LIGOScientific:2020kqk, KAGRA:2021duu, LIGOScientific:2025pvj}, but also to independently measuring cosmological parameters~\cite{LIGOScientific:2019zcs, LIGOScientific:2021aug, LIGOScientific:2025jau}.\newline\vspace{-2mm}

Although a significant amount of work has been dedicated to this task, current population analyses are still driven by modelling assumptions, especially for identifying the presence of specific features. This challenges the physical interpretation of the population as well as robust estimates of the Hubble constant. Phenomenological methods, which impose specific functional forms to model the population~\cite{Talbot:2018cva, Wysocki:2018mpo}, offer simple parameterisations but can potentially introduce systematic biases~\cite{Gennari:2025nho, Farah:2023vsc}. To mitigate this issue, the community is increasingly adopting more agnostic, non-parametric techniques, which provide flexible reconstructions that can capture unexpected features~\cite{Tiwari:2020vym, Rinaldi:2021bhm, Sadiq:2021fin, Edelman:2022ydv, Callister:2023tgi, Ray:2024hos, Heinzel:2024jlc}.
%
Since the first \ac{GW} catalog~\cite{LIGOScientific:2020kqk}, the primary mass distribution has been known to follow a decreasing, powerlaw–like behavior~\cite{Talbot:2018cva, LIGOScientific:2020kqk}, with additional structure emerging as the number of detected events has grown. Current data robustly support a high-probability peak at $10 M_{\odot}$ and an overdensity in the $30$-$40 M_{\odot}$ range~\cite{LIGOScientific:2025pvj, Sadiq:2025vly, Mould:2026sww}, along with possible features near $\sim 20 M_{\odot}$, potentially preceded by a gap~\cite{Tiwari:2025oah, Tagliazucchi:2026gxn, Bertheas:2026odj}, and a higher-mass overdensity around $60$-$70 M_{\odot}$~\cite{MaganaHernandez:2024qkz, Pierra:2026ffj, Bertheas:2026odj}. It is far from trivial to assess the statistical significance of these features. Simulated catalogs can quantify the impact of specific population realisations~\cite{Farah:2023vsc, Gennari:2025nho}, but such analyses depend on the underlying unknown distribution. Population-synthesis simulations are increasingly bridging the gap between the theory of formation channels and observations~\cite{Mahapatra:2022ngs, Golomb:2023vxm, Vaccaro:2023cwr, Colloms:2025hib, Scarpa:2026piy}, but they remain ineffective today~\cite{Rastello:2021gvw, Santoliquido:2023wzn, Sgalletta:2024jhw}. Furthermore, the results can be strongly influenced by modelling and data analysis assumptions, highlighting the need for new approaches to robustly estimate uncertainties within current methods~\cite{Rinaldi:2025emt, Rinaldi:2025evs, Corelli:2026thw}.\newline\vspace{-2mm}

In this paper, we employ weakly modelled techniques to characterise the emergence of structure in the primary mass distribution by varying the number of components in a B-spline expansion~\cite{Edelman:2021zkw, Edelman:2022ydv, Tagliazucchi:2026gxn}. Using the latest \ac{GW} catalog, GWTC-4.0, we confirm enriched low-mass structure and a non–powerlaw high-mass tail~\cite{MaganaHernandez:2024qkz, Tagliazucchi:2026gxn, Pierra:2026ffj, Bertheas:2026odj}. We find that a logarithmic spacing of the spline knots is preferred over a uniform one, suggesting a logarithmic distribution of features in the mass spectrum. These observations could be qualitatively explained by repeated hierarchical mergers. Our results are consistent with a recently proposed parametric model using tapered powerlaw components~\cite{Bertheas:2026odj}, offering a potential avenue to further test this hypothesis.

We apply our method to population-based cosmology~\cite{Taylor:2011fs, Ezquiaga:2022zkx}, the only informative approach to estimate the Hubble constant using \ac{GW} events without electromagnetic information~\cite{LIGOScientific:2025jau}.
We demonstrate that $H_0$ estimates are highly sensitive to the resolvability of features in the primary mass distribution, particularly at its boundaries, consistent with recent findings~\cite{Mali:2024wpq, MaganaHernandez:2024qkz, Tagliazucchi:2026gxn, Pierra:2026ffj, Bertheas:2026odj}. As shown in~\cite{Tagliazucchi:2026gxn}, we find that increasing the model complexity improves constraints on $H_0$. Our logarithmic-spaced 10-spline (14-spline) analysis yields an $H_0$ measurement of $H_0=64.20^{+18.75}_{-13.36}\; \text{km}\, \text{s}^{-1}\, \text{Mpc}^{-1}$ ($H_0=68.40^{+17.57}_{-12.47}\; \text{km}\, \text{s}^{-1}\, \text{Mpc}^{-1}$) with $25\%$ ($22\%$) precision at $68\%$ \ac{CI}. Analyses with larger number of splines are possibly affected by modelling bias at higher masses.

To mitigate potential systematics, we introduce a novel approach that isolates events near the $10 M_{\odot}$ peak, treating them as a coherent subpopulation. Across different population models, the reconstructed feature is consistent with low-mass results from the full \ac{BBH} population. Remarkably, only 24 events suffice to achieve a $40\%$ constraint on $H_0$, comparable to full-population analyses by the \ac{LVK} collaboration~\cite{LIGOScientific:2025jau}. We find that the low-redshift rate of $10 M_{\odot}$ events declines more steeply than that of higher-mass events, with a powerlaw index $\gamma = 8.45^{+2.69}_{-2.79}$. Furthermore, the intrinsic mass-evolution of this subset is consistent with stationarity in redshift, underscoring the potential of this strategy to reduce modelling uncertainties in cosmological inference.\newline\vspace{-2mm}

Our results have important implications for \ac{BH} astrophysics and \ac{GW} cosmology, providing a robust framework to quantify uncertainties in population analyses, reveal emergent astrophysical features, and mitigate modelling systematics for future precision cosmology.

\section{Dataset and method}
\label{sec:data_method}

We consider the \ac{BBH} events in the latest \ac{LVK}~\cite{LIGOScientific:2014pky, VIRGO:2014yos, KAGRA:2020tym, KAGRA:2013rdx} \ac{GW} transient catalog, GWTC-4.0~\cite{LIGOScientific:2025slb}, selecting those with a \ac{FAR} below one per year~\cite{LIGOScientific:2025pvj}. We exclude the two highest-mass systems, \GW{190521}~\cite{LIGOScientific:2020iuh, LIGOScientific:2020ufj} and \GW{231123}~\cite{LIGOScientific:2025rsn}, as well as one highly asymmetric binary~\cite{LIGOScientific:2020stg}, leaving a total of 150 events. These are treated as outliers; however, their inclusion is not expected to affect our conclusions, consistent with previous parametric analyses~\cite{Bertheas:2026odj, Mould:2026sww}.
We adopt a hierarchical Bayesian framework to reconstruct the intrinsic population distribution in the presence of selection effects~\cite{Gaebel:2018poe, Mandel:2018mve, Vitale:2020aaz}, using \texttt{ICAROGW}~\cite{Mastrogiovanni:2022hil, Mastrogiovanni:2023emh, Mastrogiovanni:2023zbw} to simultaneously infer astrophysical and cosmological parameters. We parametrise the primary mass, mass ratio and redshift evolution, assuming no population-level correlations between them. The mass ratio is modelled with a truncated Gaussian distribution, while the redshift rate follows a Madau–Dickinson function~\cite{Madau:2014bja}. Further details on the functional forms are provided in~\cite{Gennari:2025nho, Bertheas:2026odj}. We neglect spins in the population analysis, as no significant correlations are supported by current data~\cite{Tiwari:2025oah, Sadiq:2025vly, Tong:2025xvd} (although see~\cite{Galaudage:2021rkt, Callister:2021fpo, Biscoveanu:2022qac, Pierra:2024fbl, Berti:2025usa}).\newline\vspace{-2mm}

We adopt an agnostic modelling of the primary mass distribution using a B-spline expansion, varying the number of basis components to increase the model flexibility~\cite{Edelman:2021zkw, Edelman:2022ydv, Tagliazucchi:2026gxn}. Full implementation details are provided in App.~\ref{app:splines}. Below we highlight the key properties of our approach.
\begin{itemize}
    \item The knot vector spans the minimum and maximum masses of the distribution, which are inferred in the analysis, effectively allowing node positions to vary during inference. We consider two knot configurations, uniform spacing and logarithmic spacing, to explore how the model's degrees of freedom are distributed across the mass spectrum.
    \item We model the logarithm of the mass distribution and allow for negative spline coefficients in order to capture potential sharp features. This differs from other approaches in the literature, which impose priors on the coefficients to limit abrupt changes between neighboring components~\cite{Edelman:2022ydv, LIGOScientific:2025pvj}.
\end{itemize}
We report results using quadratic polynomials, noting that cubic splines yield similar outcomes. In App.~\ref{app:splines}, we also present results obtained by modelling the primary distribution directly with a B-spline, rather than its logarithm.\newline\vspace{-2mm}

We use the \texttt{nessai} sampler~\cite{nessai, Williams:2021qyt, Williams:2023ppp} to draw samples from the hierarchical likelihood, employing 2000 live points. For each event, we use 3000 posterior samples and incorporate the GWTC-4.0 cumulative sensitivity estimates to account for selection effects~\cite{GWTC-4-0:injections, Essick:2025zed}. To ensure numerical stability during the Monte Carlo integration, we set the likelihood to zero when the effective number of posterior samples per event is below 10 or when the effective number of injections is less than four times the number of events~\cite{Farr:2019rap, KAGRA:2021duu} (further discussion in App.~\ref{app:numerical_stability}).

\section{Astrophysical distribution}
\label{sec:population}

In this section, we fix the cosmology to the fiducial values of Ref.~\cite{Planck:2015fie} and focus on inferring the properties of the astrophysical population. We repeat the analysis while varying the number of spline basis elements, allowing us to resolve the emergence of structure in the primary mass distribution as a function of the model resolution.\newline\vspace{-2mm}

\begin{figure}
    \centering

    \begin{subfigure}{\linewidth}
        \centering
        \includegraphics[width=\linewidth]{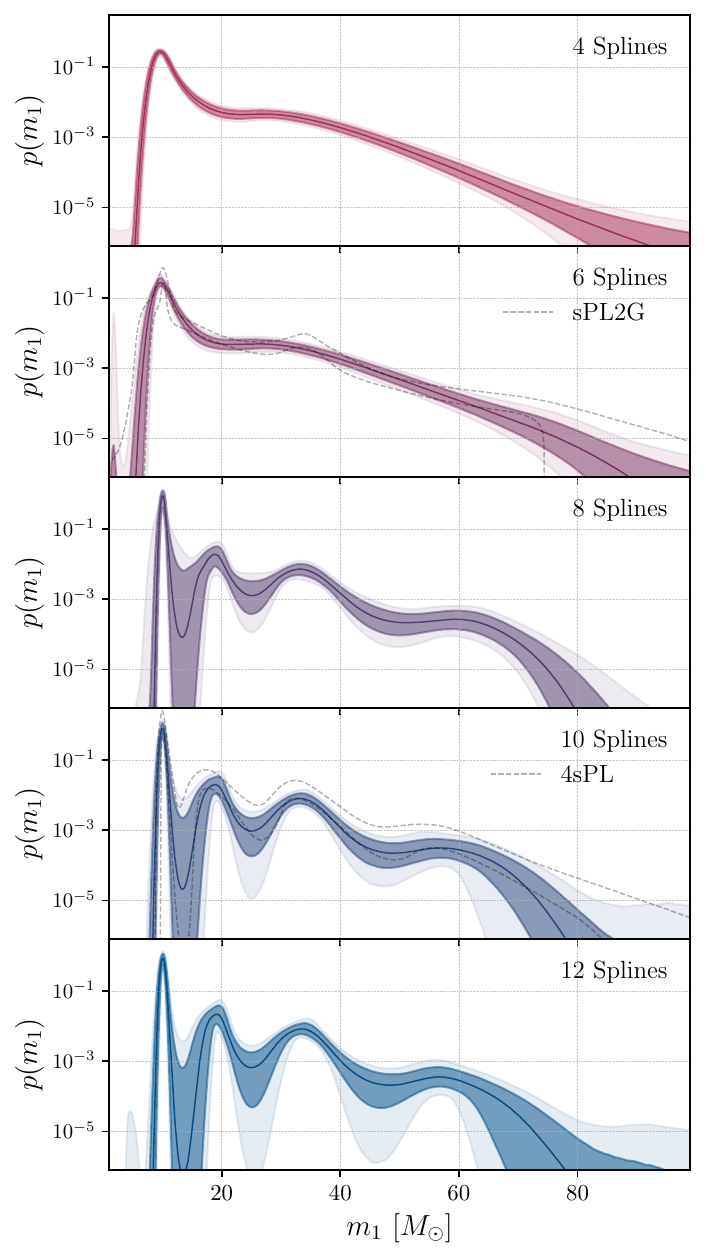}
    \end{subfigure}

    \begin{subfigure}{\linewidth}
        \centering
        \includegraphics[width=\linewidth]{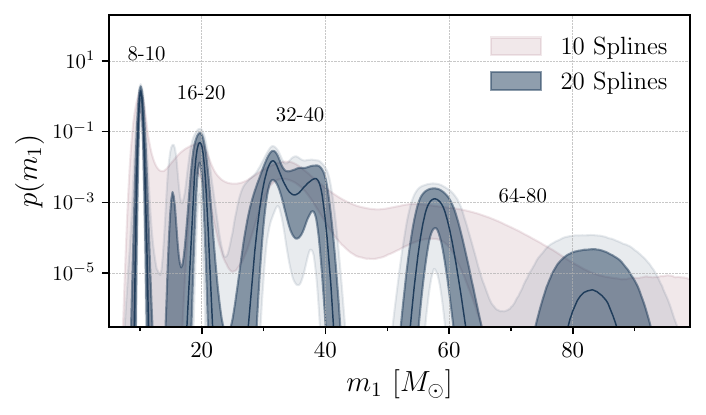}
    \end{subfigure}

    \caption{\justifying \footnotesize Reconstructed primary mass distributions for different logarithmically spaced spline models on GWTC-4.0 data, showing the median, $68\%$ and $90\%$ \acp{CI}. The dashed curves indicate the $68\%$ \ac{CI} from the phenomenological analysis of~\cite{Bertheas:2026odj}, based on a tapered powerlaw with two Gaussians (sPL2G) and four tapered powerlaws (4sPL). In the final panel, the intervals associated with each feature correspond to twice the mass range of the preceding feature.}
    \label{fig:GWTC-4.0_pop_spline}
\end{figure}

Fig.~\ref{fig:GWTC-4.0_pop_spline} shows the inferred population for a logarithmic spacing of the knot vector. A small number of components yields a smoothed reconstruction, while more splines allow increasingly complex structure to be resolved. Because the support of each spline element depends on the total number of components, models with more splines have greater flexibility on shorter scales, whereas models with fewer splines are constrained by the node positions and can only capture broader features.
%
Analyses with 4 and 6 splines recover a powerlaw–like mass spectrum with only a broad overdensity near $\sim 30 M_{\odot}$. From 8 splines onward, additional structure becomes apparent, including a sharp peak at $10 M_{\odot}$, a possible underdensity near $\sim 15 M_{\odot}$, and three other overdensities around $20 M_{\odot}$, $30$-$40 M_{\odot}$, $50$-$70 M_{\odot}$. These features remain stable as the number of splines is further increased, with additional components primarily enhancing the resolution of this structure.

The presence of these features has been suggested in previous studies, in addition to the robust global maximum at $10 M_{\odot}$ and the peak at $35 M_{\odot}$~\cite{Talbot:2018cva, LIGOScientific:2020kqk, Callister:2023tgi, Mould:2026sww}. Several analyses have identified a pronounced $10 M_{\odot}$ peak, present even in earlier catalogs~\cite{Toubiana:2023egi, Gennari:2025nho, Bertheas:2025mzd}. A $20 M_{\odot}$ overdensity, potentially preceded by a gap, has also been discussed since GWTC-3~\cite{Toubiana:2023egi, Tiwari:2025lit, Gennari:2025nho}, and our analysis confirms that current data support its presence~\cite{Legred:2026oiz}, though the gap remains statistically uncertain due to large reconstruction errors. We further identify a non–powerlaw high-mass tail, consistent with previous findings~\cite{LIGOScientific:2025pvj, MaganaHernandez:2024qkz, Pierra:2026ffj}. Recently, Ref.~\cite{Bertheas:2026odj} proposed a phenomenological parametrisation based on four tapered powerlaw components, capable of simultaneously fitting this structure. We support this result using our more agnostic modelling approach. In Fig.~\ref{fig:GWTC-4.0_pop_spline}, we compare their reconstruction with our 10-spline analysis (fourth panel), while the second panel shows their powerlaw plus two Gaussians model against a 6-spline reconstruction. These results highlight that limited model flexibility can hinder the recovery of this structure, which remains consistently supported by current \ac{GW} data despite the increased sample size.\newline\vspace{-2mm}

Conversely, overly flexible models may suffer from overfitting. Tab.~\ref{tab:BFs} in Appendix reports the \acp{BF} for the different analyses, showing that the evidence generally increases as more components are added, despite the additional parameters. The fact that the gain in maximum likelihood results in only moderate \acp{BF} indicates that the extra parameters are not tightly constrained, as reflected by strong correlations and residual model degeneracies in the posteriors. Moreover, numerical-stability cuts complicate the interpretation of these values, as they exclude a significant portion of the parameter space where the likelihood may exhibit non-trivial behavior~\cite{Mould:2025dts, Bertheas:2026odj}. As a result, the \ac{BF} may not provide a fully reliable figure of merit in this setup. While we cannot rule out that the data genuinely favor models that enhance clustering of events, further tests on simulated catalogs will be necessary to assess the robustness of this conclusion.\newline\vspace{-2mm}

\begin{figure*}
    \centering
    \includegraphics[width=\linewidth]{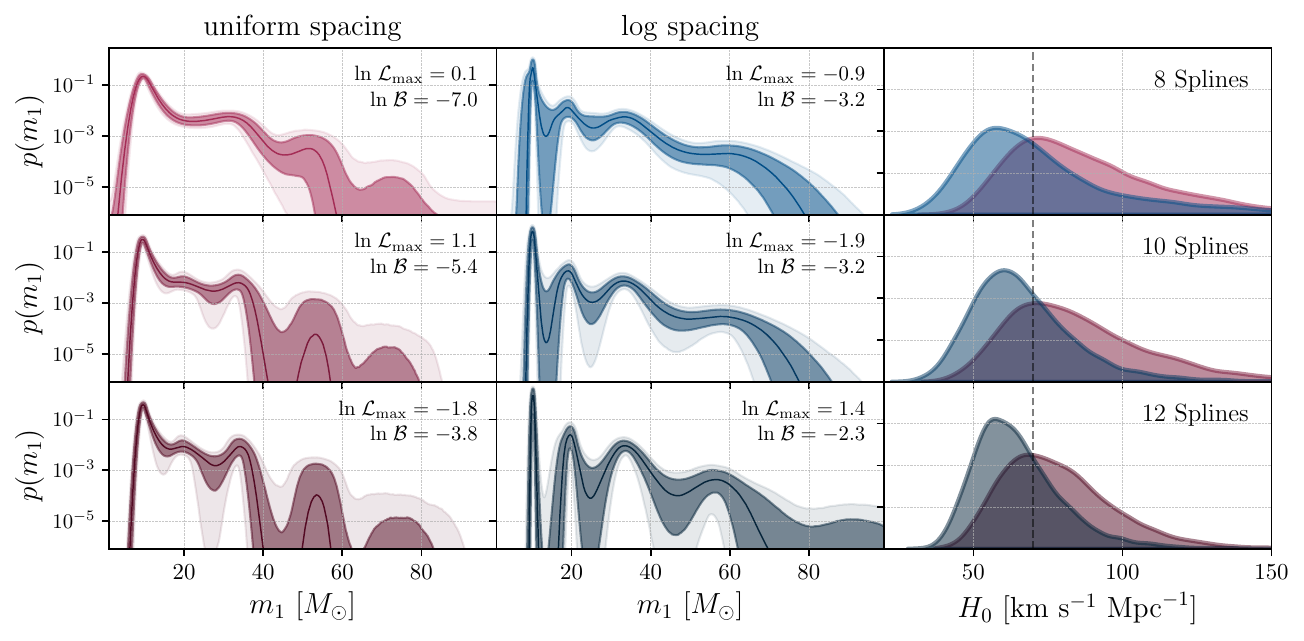}
    \caption{\justifying \footnotesize Reconstructed primary mass distributions for spline models with uniform (\textit{left}) and logarithmic (\textit{center}) knot spacing on GWTC-4.0 data, showing the median, the $68\%$ and $90\%$ \acp{CI}. We report the natural logarithm of the maximum-likelihood values and the \acp{BF} relative to the 10-spline analysis with logarithmic knot spacing and fixed cosmology. The \textit{right} column shows the corresponding $H_0$ posteriors, with the dashed vertical line indicating 70.}
    \label{fig:GWTC-4.0_spline_cosmology}
\end{figure*}

We argue that the observed distribution of features could be qualitatively explained by repeated hierarchical mergers~\cite{Mapelli:2021syv, Gerosa:2021mno}. Fig.~\ref{fig:GWTC-4.0_pop_spline} shows the 20-spline analysis, revealing distinct clusters of events corresponding to consecutive merger generations. The mass of these components increases with successive generations while their probability decreases, consistent with a population of mergers having mass ratios in (0.8, 1) that undergo repeated coalescences. Several studies have suggested signatures of hierarchical mergers in current data~\cite{Fishbach:2017dwv, Kimball:2020opk, Tagawa:2021ofj, Mould:2022ccw, Li:2023yyt, Banagiri:2025dmy}, some of which align with this picture~\cite{Tiwari:2020otp, Kimball:2020qyd, Godfrey:2023oxb, Tiwari:2025oah, Ginat:2026awh}. This result does not exclude the possibility of having several formation channels for different portions of the mass spectrum~\cite{Li:2023yyt, Godfrey:2023oxb}. Further analyses will be needed to better investigate possible population-level correlations within the low-mass spectrum in order to investigate this hypothesis, although more information from future data will probably be needed. We note that the spiky features in our reconstruction become less pronounced under different prior choices~\cite{LIGOScientific:2025pvj, Tagliazucchi:2026gxn} or when using alternative agnostic parametrisations~\cite{Rinaldi:2023bbd, Farah:2023vsc, Tiwari:2025oah}.
Finally, we note that adopting a uniform spacing for the spline nodes yields markedly different reconstructions, in which sharp low-mass structure is no longer resolved and variability increases at high masses (results not shown, but similar to those with a free cosmology reported in Sec.~\ref{sec:cosmology}). This parametrisation is disfavored relative to the logarithmic spacing, with \acp{BF} ranging from $10^{2.2}$ to $10^{1.2}$ for 6 to 12 splines, suggesting that the data prefer logarithmically spaced nodes. This further supports the view that the mass spectrum is naturally distributed in logarithmic space.

\section{Cosmological inference}
\label{sec:cosmology}

Population inference also enables the measurement of cosmological parameters under the assumption that the intrinsic distribution exhibits negligible redshift evolution~\cite{Markovic:1993cr, Chernoff:1993th, Finn:1995ah, Taylor:2011fs, Ezquiaga:2022zkx}. In this framework, mass scales in the source-frame population encode redshift information, and the presence of structure in the distribution can enhance cosmological constraints. We extend the analysis of Sec.~\ref{sec:population} by jointly inferring population parameters together with the Hubble constant $H_0$ and present-day matter density $\Omega_{m,0}$ within a fiducial $\Lambda$CDM cosmological model~\cite{SupernovaSearchTeam:1998fmf, SupernovaCosmologyProject:1998vns}.\newline\vspace{-2mm}

\begin{figure*}[t]
    \centering
    \includegraphics[width=\linewidth]{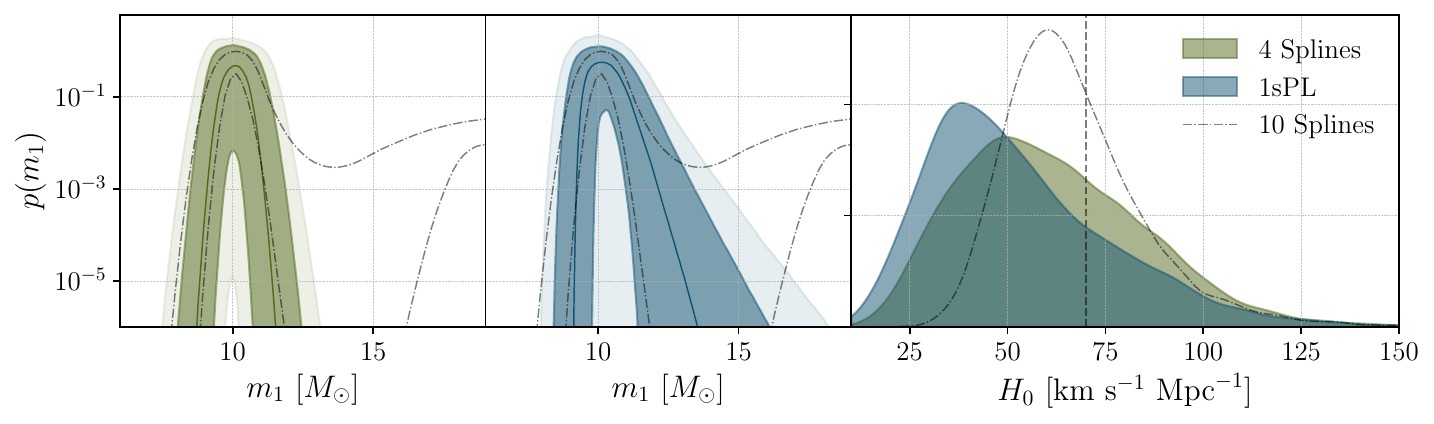}
    \caption{\justifying \footnotesize Reconstructed primary mass distributions using 24 events from GWTC-4.0 around the $10M_{\odot}$ peak, modeled with a uniform-knot 4-spline (\textit{left}) and a tapered powerlaw (\textit{center}), showing the median, $68\%$ and $90\%$ \acp{CI}. The dash-dotted curves correspond to the logarithmically spaced 10-spline analysis of the full \ac{BBH} population. The \textit{right} column shows the corresponding $H_0$ posteriors, with the dashed vertical line indicating 70.}
    \label{fig:GWTC-4.0_10M-only}
\end{figure*}

Fig.~\ref{fig:GWTC-4.0_spline_cosmology} shows the reconstructed distributions for the two knot spacings and their corresponding posteriors on $H_0$. Overall, constraints improve as the number of spline components increases. Logarithmic spacing yields more precise measurements, driven by sharper low-mass features. For 10 and 12 splines, we obtain $H_0=64.20^{+18.75}_{-13.36}\; \text{km}\, \text{s}^{-1}\, \text{Mpc}^{-1}$ ($H_0=81.19^{+28.41}_{-18.39}\; \text{km}\, \text{s}^{-1}\, \text{Mpc}^{-1}$) and $H_0=62.66^{+16.22}_{-11.24}\; \text{km}\, \text{s}^{-1}\, \text{Mpc}^{-1}$ ($H_0=75.93^{+21.13}_{-15.43}\; \text{km}\, \text{s}^{-1}\, \text{Mpc}^{-1}$) for logarithmic (uniform) spacing, corresponding to precisions of $25\%$ ($29\%$) and $22\%$ ($24\%$) at $68\%$ \ac{CI} respectively. With logarithmic spacing, 14 splines yield $H_0=68.40^{+17.57}_{-12.47}$, corresponding to a $22\%$ constraint. For 16 splines, the posterior shifts toward lower values, $H_0=50.26^{+16.30}_{-8.00}$, and the reconstructed distribution approaches that shown in the bottom panel of Fig.~\ref{fig:GWTC-4.0_pop_spline}. A similar shift has been reported in other analyses, both with phenomenological~\cite{Bertheas:2026odj, Mali:2024wpq} and more agnostic approaches~\cite{Tagliazucchi:2026gxn}. For 18 and 20 splines, the reconstructed distribution differs significantly from the corresponding fixed-cosmology analysis, and the $H_0$ posteriors shift to very high values, $\sim 180\, \text{km}\, \text{s}^{-1}\, \text{Mpc}^{-1}$. This suggests that the additional degrees of freedom introduced by cosmological parameters are effectively used to compress the population reconstruction, clustering events into a small number of sharp peaks (see Fig.~\ref{fig:gwtc-4_20-spline_cosmology} in Appendix). This highlights how insufficiently controlled population modelling can lead to results that are difficult to physically interpret, and different prior assumptions may alleviate these issues~\cite{Tagliazucchi:2026gxn}. On the other hand, uniform spacing leads to more consistent results when the number of splines is increased.
The measurements on $\Omega_{m,0}$ are in general uninformative, only marginally excluding limited regions of the parameter space. Logarithmic spacing tends to disfavour values close to zero, while uniform spacing values near unity.

We also present in Fig.~\ref{fig:gwtc-4_spline-PDF_cosmology} the same analysis modelling the primary mass distribution directly with a spline expansion, rather than its logarithm (see App.~\ref{app:splines}). While this choice reduces the effective resolution for a fixed number of components, it produces reconstructions with sharper boundaries that still yield competitive constraints on $H_0$ (although being disfavoured by evidence, see App.~\ref{app:additional_results}). These results show that both the precision and location of the $H_0$ posterior are primarily driven by the sharpness of the low- and high-mass features: a sharper high-mass cutoff shifts the posterior toward $\sim 70\; \text{km}\, \text{s}^{-1}\, \text{Mpc}^{-1}$, while a more pronounced $10 M_{\odot}$ peak tightens the constraints. This interpretation is consistent with previous studies showing that improved characterisation of the high-mass~\cite{MaganaHernandez:2024qkz, Pierra:2026ffj} and low-mass~\cite{Tagliazucchi:2026gxn, Bertheas:2026odj} structure provides additional constraining power. Our results offer a unified set of reconstructions that capture both effects simultaneously.

More broadly, this highlights the strong sensitivity of cosmological inference to population-modelling assumptions, which govern the structure of the mass distribution and, in turn, impact both the width and location of the $H_0$ posterior. We argue that this is particularly problematic in the high-mass regime, where the smaller number of events means these effects are more significant~\cite{Fishbach:2019ckx, Mould:2026sww, Gennari:2025nho}.

\section{Subpopulation analysis}
\label{sec:subpop_cosmology}

The large uncertainties in the primary mass distribution and their impact on cosmological inference highlight the need for more robust approaches for population-based cosmology. We propose a complementary strategy that isolates a subpopulation of events around the $10 M_{\odot}$ peak, potentially sharing common properties and thus less prone to modelling systematics. This is particularly advantageous compared to the high-mass regime, where sparse observations over a broad mass and redshift range increase the risk of bias from unmodelled effects, such as redshift evolution or other correlations~\cite{Mukherjee:2021rtw, Pierra:2023deu, Agarwal:2024hld, Gennari:2025nho, Tong:2025xvd}. Motivated by the results in Sec.~\ref{sec:population} and the potential presence of a low-mass gap, we select 24 events around the $10M_{\odot}$ peak with median source-frame primary masses below $13M_{\odot}$, leveraging the posterior separation in analogy with standard treatments of neutron star populations~\cite{KAGRA:2021duu, LIGOScientific:2025pvj} (although this criterion may be affected by event misclassification~\cite{Fishbach:2019ckx, Galaudage:2019jdx}). Additional tests reported in App.~\ref{app:sub-population_supplemental} have been used to benchmark our method on simulated data.\newline\vspace{-2mm}

In Fig.~\ref{fig:GWTC-4.0_10M-only} we show the reconstructed distribution of the $10M_{\odot}$ subpopulation and the corresponding $H_0$ posteriors. Results are shown for both a uniform knot-spaced 4-spline analysis, providing a flexible parametrisation, and a single tapered powerlaw model. Both approaches yield similar reconstructions, in close agreement with the 10-spline analysis of the full catalog at low mass. Remarkably, constraints on $H_0$ using only 24 events are comparable to those obtained with the full set of 150 events, demonstrating that the $10M_{\odot}$ feature alone carries substantial information on the Hubble constant. We obtain $H_0=58.60^{+26.92}_{-20.85}\; \text{km}\, \text{s}^{-1}\, \text{Mpc}^{-1}$ ($H_0=48.35^{+28.66}_{-17.16}\; \text{km}\, \text{s}^{-1}\, \text{Mpc}^{-1}$) using the 4-spline (powerlaw) model, therefore a $40\%$ ($48\%$) precision at $68\%$ \ac{CI}, comparable to the $44\%$ reported by \ac{LVK} for the full \ac{BBH} population~\cite{LIGOScientific:2025jau}. The $H_0$ posterior from the single powerlaw model is slightly shifted toward lower values, consistent with results from powerlaw–only models applied to the full population~\cite{Bertheas:2026odj} and with our analysis in Sec.~\ref{sec:cosmology}. We verified that the results are largely insensitive to the number of spline elements, as well as to using cubic instead of quadratic splines. Similarly, a single truncated Gaussian yields consistent findings. Compared to the spline analysis, parametric models are more sensitive to the two lowest-mass events (see Fig.~\ref{fig:10M_posteriors_KDE}), which introduce multimodalities in the inferred minimum mass. To mitigate this, we truncate the lower prior to $m_{min}=7 M_{\odot}$ in the powerlaw analysis, isolating the main mode consistent with the spline results and ensuring the posteriors remain well-peaked away from the prior bounds. Future work will explore whether alternative parametrisations yield improved modelling.\newline\vspace{-2mm}

The mass ratio distribution inferred from the subpopulation is consistent with that of the full population, albeit less well constrained (see Fig.~\ref{fig:GWTC-4_q-z}). In contrast, the inferred redshift rate evolution differs significantly. In the 10-spline analysis, the low-redshift evolution exponent is recovered as $\gamma = 3.73^{+0.66}_{-0.60}$ (free cosmology) and $\gamma = 3.62^{+0.66}_{-0.61}$ (cosmology fixed to~\cite{Planck:2015fie}), consistent with the literature~\cite{LIGOScientific:2025pvj, LIGOScientific:2025jau}. For the $10M_{\odot}$ subpopulation, however, we find significantly larger values, $\gamma = 11.63^{+7.19}_{-4.74}$ and $\gamma = 8.45^{+2.69}_{-2.79}$, respectively. This difference indicates that the assumption of uniform \ac{BBH} properties across the full population may be oversimplified, as low-mass events appear to decrease in rate faster than high-mass events over time. If the hierarchical-merger hypothesis discussed in Sec.~\ref{sec:population} is correct, the distinct rate evolution at higher masses could, in principle, provide constraints on the timescales of repeated mergers.

Finally, we investigate the presence of intrinsic redshift evolution in the population. This is particularly relevant for cosmological inference, as unaccounted mass–redshift correlations could introduce significant biases~\cite{Mukherjee:2021rtw, Pierra:2023deu, Agarwal:2024hld, Mali:2024wpq, Gennari:2025nho, Tong:2025xvd}. We adopt a non-tapered, four-powerlaw model in which the characteristic scale of the first feature evolves linearly with redshift~\cite{Gennari:2025nho}, assuming a fixed cosmology~\cite{Planck:2015fie}. Modelling the minimum mass as $m_{min}(z)=m_{min,0} + \alpha\, z$, we obtain $\alpha = -0.34^{+2.90}_{-1.89}$, consistent with no evolution. This result agrees with previous studies~\cite{Karathanasis:2022rtr, Ray:2023upk, Rinaldi:2023bbd, Heinzel:2024hva, Sadiq:2025aog, Lalleman:2025xcs, Gennari:2025nho}, while providing informative posteriors that support a stationary population, further justifying the assumption of stationarity adopted for cosmological inference. Letting the cosmology free to vary, the posterior on $H_0$ is consistent with that obtained when no evolution is included, although a robust characterisation of the degeneracies arising when simultaneously inferring mass evolution and cosmology is beyond the scope of this work and will be essential for enabling reliable future cosmological analyses.

\begin{figure}
    \centering
    \includegraphics[width=\linewidth]{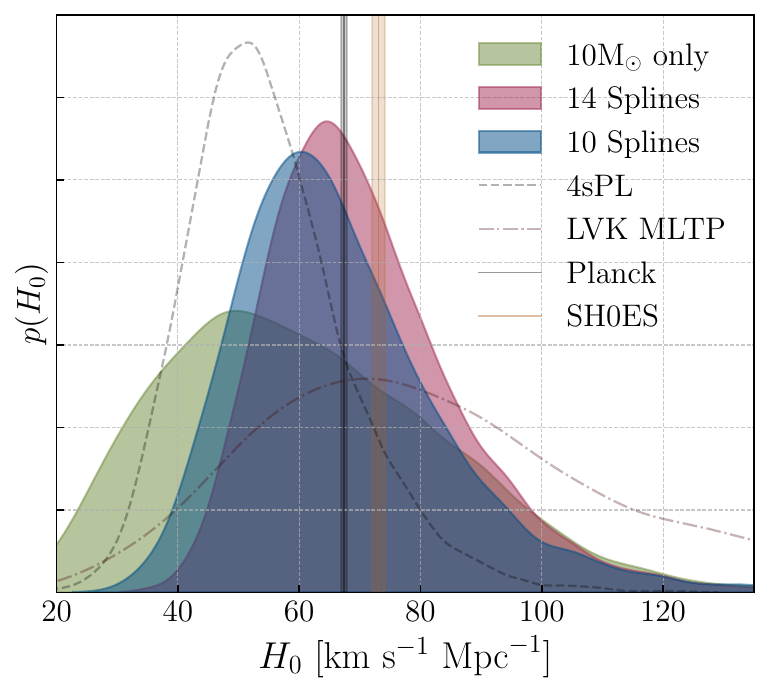}
    \caption{\justifying \footnotesize Marginal posterior distributions for the Hubble constant from GWTC-4.0 data. Solid curves show results from logarithmically spaced spline models using the full \ac{BBH} population with 10 (\textit{blue}) and 14 (\textit{pink}) components, and from a 4-spline model with uniform knots (\textit{green}) using 24 events around the $10M_{\odot}$ peak. The dashed curves correspond to a four tapered powerlaw model (4sPL) from~\cite{Bertheas:2026odj}, while the dash-dotted to a powerlaw with two Gaussians model (MLTP) from \ac{LVK}~\cite{LIGOScientific:2025jau}. Vertical bands indicate the current $68\%$ \ac{CI} constraints from early~\cite{Planck:2018vyg} and late~\cite{Riess:2021jrx} Universe measurements.}
    \label{fig:h0_posteriors}
\end{figure}

\section{Conclusions}
\label{sec:conclusions}

In this paper, we investigate the ability to reconstruct structure in the \ac{BH} mass distribution from \ac{GW} observations. Using a flexible B-spline expansion, we characterise the emergence of features in the population as the number of spline components increases. We find evidence for four overdensities in the GWTC-4.0 primary mass distribution once sufficient model flexibility is allowed, in agreement with previous studies~\cite{Toubiana:2023egi, Gennari:2025nho, Bertheas:2025mzd, Bertheas:2026odj, Tiwari:2025lit, Tiwari:2025oah, Tagliazucchi:2026gxn, Legred:2026oiz}. Our 10-spline analysis with logarithmically spaced knots closely reproduces the results of the four–powerlaw component model of Ref.~\cite{Bertheas:2026odj}, while Bayesian evidence mildly favors a logarithmic distribution of the population features. We argue that the emergence of distinct clusters in highly flexible models may point to repeated hierarchical mergers as a dominant formation channel. Testing this interpretation will require further analyses probing differences in population properties between low and higher masses.\newline\vspace{-2mm}

We extend our analysis to cosmological inference using the \ac{BBH} population alone. We obtain improved constraints on the Hubble constant compared to recent \ac{LVK} results~\cite{LIGOScientific:2025jau}, reaching a $22\%$ precision with 12–14 logarithmically spaced splines. We show that both low- and high-mass structures play a key role in shaping the $H_0$ posterior, affecting both its width and location. We further propose a complementary approach that targets events around $10M_{\odot}$ for cosmological inference. If this subpopulation shares common properties, it can mitigate modelling systematics at the cost of a reduced dataset. Remarkably, using only 24 events, we recover a mass distribution consistent with the low-mass spectrum of the full population and obtain a $40\%$ constraint on $H_0$. Fig.~\ref{fig:h0_posteriors} summarises our $H_0$ measurements across models. Finally, we find that the low-redshift rate of $10M_{\odot}$ events declines faster than that of higher-mass events, compared to full-catalog analyses, suggesting that the common assumption of a shared rate evolution may be inaccurate. We also find that intrinsic mass–redshift correlations for the $10M_{\odot}$ subpopulation are consistent with stationarity under a fixed cosmology, reinforcing the robustness of our results and showing the potential of this approach for robust cosmological inference.\newline\vspace{-2mm}

Our results demonstrate that the reconstruction of the primary mass distribution is highly sensitive to modelling choices. Parametrisations with limited flexibility may underestimate uncertainties, whereas more agnostic approaches naturally yield smoother distributions with larger, more conservative uncertainties~\cite{Ray:2023upk, Callister:2023tgi, Mould:2026sww}. Relaxing these assumptions can reveal additional structure in the current data, highlighting potential limitations of commonly used phenomenological models. Addressing these modelling systematics is crucial for both astrophysics and cosmology with \acp{GW}. Future work should develop methods to systematically compare analyses, assess robustness, and identify over- or under-fitting in population inference.

\begin{acknowledgments}

V.G. thanks the many colleagues who provided valuable feedback and helped improve this work. In particular, he acknowledges Stefano Rinaldi, Gabriele Demasi, Michela Mapelli, Alexander Papadopoulos, Jose Ezquiaga, Rico Lo, Jonathan Gair, Harald Pfeiffer, Matteo Tagliazucchi, Nicola Borghi, Thomas Dent, Sylvain Marsat, John Baker, Manuel Piarulli for insightful discussions. The authors also thank the \ac{LVK} Cosmology Working Group for useful interactions and Vaibhav Tiwari, Utkarsh Mali and Yonadav Barry Ginat for their comments. This paper was reviewed by the LIGO-Virgo-KAGRA Collaboration (document number: P2600167).\newline

\footnotesize{This project has received financial support from the CNRS through the AMORCE funding framework and from the Agence Nationale de la Recherche (ANR) through the MRSEI project ANR-24-MRS1-0009-01.
The authors acknowledge support form the French space agency CNES in the framework of LISA. T.B. acknowledges support from a CDSN PhD grant from ENS-PSL.
%
The authors are grateful for computational resources provided by the IN2P3 computing centre (CC-IN2P3) in Lyon (Villeurbanne).
%
This research made use of data, software and/or web tools obtained from the Gravitational Wave Open Science Center~\cite{LIGOScientific:2019lzm, KAGRA:2023pio}, a service of the LIGO Scientific Collaboration, the KAGRA Collaboration and the Virgo Collaboration.
LIGO Laboratory and Advanced LIGO are funded by the United States National Science Foundation (NSF) as well as the Science and Technology Facilities Council (STFC) of the United Kingdom, the Max-Planck Society (MPS), and the State of Niedersachsen/Germany for support of the construction of Advanced LIGO and construction and operation of the GEO600 detector. Additional support for Advanced LIGO was provided by the Australian Research Council. Virgo is funded, through the European Gravitational Observatory (EGO), by the French Centre National de Recherche Scientifique (CNRS), the Italian Istituto Nazionale di Fisica Nucleare (INFN) and the Dutch Nikhef, with contributions by institutions from Belgium, Germany, Greece, Hungary, Ireland, Japan, Monaco, Poland, Portugal, Spain. KAGRA is supported by Ministry of Education, Culture, Sports, Science and Technology (MEXT), Japan Society for the Promotion of Science (JSPS) in Japan; National Research Foundation (NRF) and Ministry of Science and ICT (MSIT) in Korea; Academia Sinica (AS) and National Science and Technology Council (NSTC) in Taiwan.
This material is based upon work supported by NSF's LIGO Laboratory which is a major facility fully funded by the National Science Foundation.}

\end{acknowledgments}

\appendix
\renewcommand{\thefigure}{A.\arabic{figure}} 
\setcounter{figure}{0}  
\section*{Appendix}

\section{B-splines}\label{app:splines}

\begin{figure*}
    \centering
    \includegraphics[width=\linewidth]{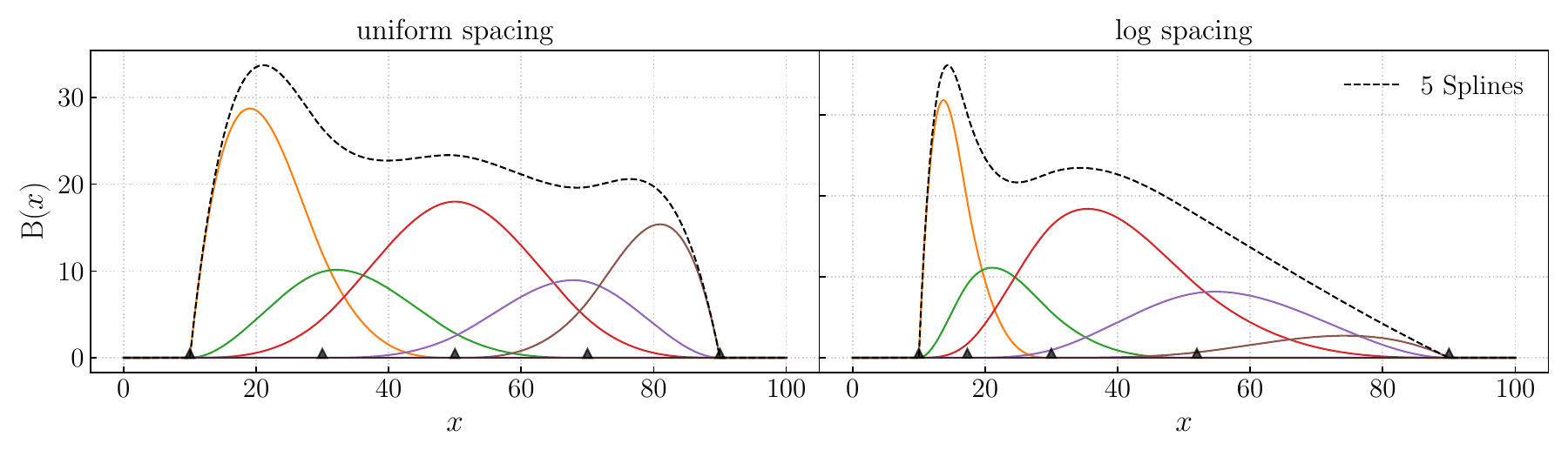}
    \vspace{-0.5cm}
    \caption{\justifying \footnotesize Representation of a cubic B-spline with five basis elements in colors, using uniform (\textit{left}) and logarithmic (\textit{right}) knot spacing, indicated by the black triangles.}
    \label{fig:B-splines}
\end{figure*}

In this section, we describe the implementation and key properties of B-splines. B-splines are smooth functions constructed from piecewise polynomials defined over a knot vector~\cite{Schoenberg:1946, Schoenberg:1973},
\begin{equation}
B(x) = \sum_{i=1}^{n} c_i\, b_i(x),
\end{equation}
where $b_i(x)$ are the spline basis elements and $c_i$ their coefficients. A spline is specified by the polynomial degree, the number of basis elements, and the knot vector. For degree $k$, each basis element has compact support over $k+1$ knot intervals, as shown in Fig.~\ref{fig:B-splines}. In our implementation, we set the first and last basis elements to zero to avoid non-zero boundary contributions. Increasing the degree broadens the support of each element, while increasing their number (at fixed bounds) reduces it. As a result, splines with many components provide a flexible representation, with localised basis functions that can capture complex structure.

For fixed $n$ and $k$, there remains significant freedom in the placement of the knot vector. In this work, we consider both uniform and logarithmic spacing between the minimum and maximum values. When modelling the primary mass distribution, uniform spacing provides an agnostic parametrisation, while logarithmic spacing allocates more resolution to low masses by concentrating degrees of freedom in that region. During inference, the minimum and maximum bounds are treated as free parameters, so that the knot vector is effectively updated at each likelihood evaluation.\newline\vspace{-2mm}

B-splines are commonly constructed via the Cox–de Boor recursion~\cite{Cox:1972, deBoor:1972}, which generates basis functions of arbitrary degree. For quadratic and cubic splines, the recursion admits explicit analytic expressions. On each knot span, the basis elements can be expressed in terms of Bernstein polynomials, enabling efficient evaluation and analytic integration for normalisation~\cite{PieglTiller1997, Farin1996}. This is particularly convenient when modelling a \ac{PDF},
\begin{equation}
p(x) = \frac{B(x)}{\int B(x)\, \dd x}
= \frac{\sum_{i=1}^{n} c_i\, b_i(x)}{\sum_{i=1}^{n} c_i \int b_i(x)\, \dd x}.
\end{equation}
This parametrisation is invariant under a global rescaling of the coefficients, $\vb{c}\rightarrow a\,\vb{c}$, since B-splines form a partition of unity. This introduces a redundant (gauge) degree of freedom, i.e. a redundant dimensionality of the model from which the observations are independent, which must be fixed. In our framework, this is removed by setting the first and last elements of the spline basis to zero. Additionally, we note that imposing Gamma priors on the spline coefficients induces a Dirichlet distribution on the simplex, effectively providing a uniform prior over the functional space spanned by the basis elements.

An alternative parametrisation models the logarithm of the \ac{PDF},
\begin{equation}
p(x) = \frac{e^{B(x)}}{\int e^{B(x)}\, \dd x}
= \frac{\exp\left(\sum_{i=1}^{n} c_i\, b_i(x)\right)}{\int \exp\left(\sum_{i=1}^{n} c_i\, b_i(x)\right)\, \dd x}.
\end{equation}
In this case, the analytic normalisation is no longer available, and global shifts of the coefficients no longer correspond to a gauge transformation. This introduces near-degeneracies that make the parameter space more challenging to sample. The key advantage is that coefficients can be negative while the exponential mapping ensures a positive-definite \ac{PDF}, allowing the reconstruction of sharper features that are difficult to capture when modelling the \ac{PDF} directly. For these reasons, we adopt logarithmic-\ac{PDF} modelling throughout this work, with a uniform prior on the coefficients, and compare results with the other approach in App.~\ref{app:additional_results}.

\section{GWTC-4.0 analysis}\label{app:additional_results}

\begin{figure*}
    \centering
    \includegraphics[width=\linewidth]{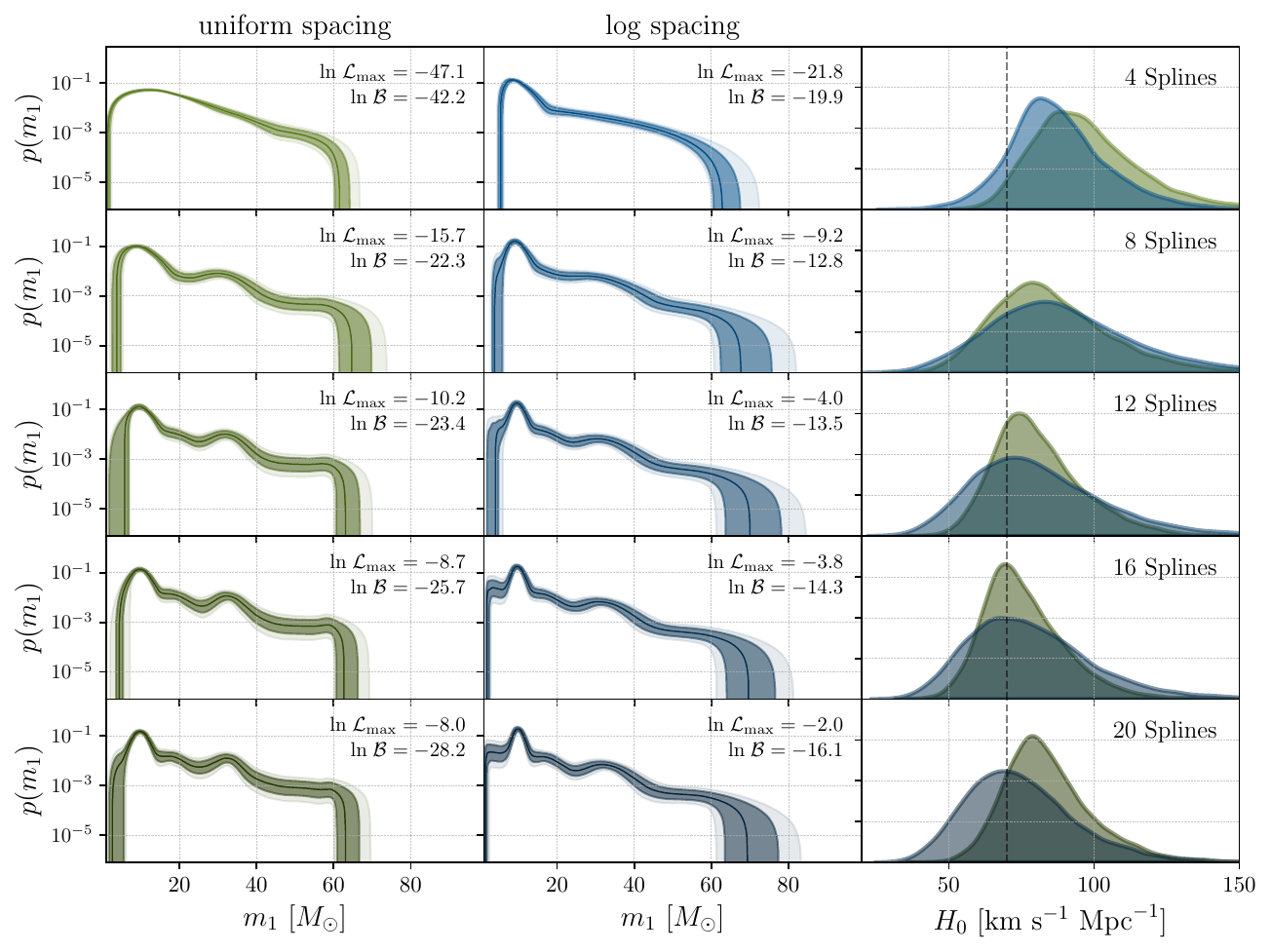}
    \caption{\justifying \footnotesize Reconstructed primary mass distributions for spline models with uniform (\textit{left}) and logarithmic (\textit{center}) knot spacing on GWTC-4.0 data, showing the median, the $68\%$ and $90\%$ \acp{CI}. In this analysis, the spline expansion directly parametrises the primary mass \ac{PDF}, rather than its logarithm. We report the natural logarithm of the maximum-likelihood values and the \acp{BF} relative to the 10-spline analysis with logarithmic knot spacing and fixed cosmology. The \textit{right} column shows the corresponding $H_0$ posteriors, with the dashed vertical line indicating 70.}
    \label{fig:gwtc-4_spline-PDF_cosmology}
\end{figure*}

We report here additional results on the GWTC-4.0 analysis using the full \ac{BBH} population. In Fig.~\ref{fig:gwtc-4_spline-PDF_cosmology} we present the analysis obtained by modelling the primary mass distribution directly with B-splines, rather than its logarithm as in the main text. The implementation is described in App.~\ref{app:splines}. This choice effectively defines a different population model, primarily in the allowed width and shape of the features. In particular, a quadratic spline expansion of the logarithm of the \ac{PDF} behaves similarly to a mixture of localised components, whereas directly modelling the \ac{PDF} constrains each basis element to follow a broader quadratic profile. As a result, the reconstructed distribution is generally smoother across the population support. At the same time, we observe sharper edges and a low-mass tail extending to smaller values, partly driven by the different scaling of the distribution tails. Finally, we emphasise that, in addition to the change in parametrisation, this approach adopts a different prior on the spline coefficients, corresponding to a uniform coverage of the distribution space. We note that this parametrisation is strongly disfavoured by Bayesian evidence, as illustrated in Fig.~\ref{fig:gwtc-4_spline-PDF_cosmology}, with the Bayes factor for the preferred 8-spline model being $\sim 10^{-5}$. Similarly, logarithmic knot spacing is strongly preferred over uniform spacing by several orders of magnitude.

\begin{table}
    \centering
    \renewcommand{\arraystretch}{1.} 
    \sisetup{table-format = -2., table-number-alignment = center}
    \begin{tabular}{c @{\hspace{0.9cm}} S @{\hspace{0.8cm}} S @{\hspace{1.cm}} S S}
    \hline
    & \multicolumn{2}{c}{\textbf{\hspace{-1mm}Fixed cosmology}} 
    & \multicolumn{2}{c}{\textbf{Cosmology}} \\[-0.6mm]
    \cmidrule(lr){2-5}
    $\#$ 
    & \multicolumn{1}{l}{$\:\:\:\ln \mathcal{B}$} 
    & \multicolumn{1}{l}{$\ln \mathcal{L}_{\max}$} 
    & \multicolumn{1}{r}{$\quad\:\:\ln \mathcal{B}$} 
    & \multicolumn{1}{r}{$\quad\ln \mathcal{L}_{\max}$} \\
    \hline
    2  & -23.2 & -32.3 & -23.0 & -30.6 \\
    4  & -1.1  & -5.5  & -2.1  & -5.2  \\
    6  & -3.0  & -6.3  & -3.4  & -7.3  \\
    8  & -1.7  & -1.3  & -3.2  & -0.9  \\
    10 & 0.0   & 0.0   & -3.2  & -1.9  \\
    12 & 0.2   & -0.6  & -2.3  & 1.4   \\
    14 & 1.5   & 2.4   & -2.8  & 0.0   \\
    16 & 2.1   & 3.0   & -3.6  & 7.6   \\
    18 & -0.5  & -0.6  & -8.3  & -11.9 \\
    20 & 3.6   & 4.0   & -9.6  & -15.4 \\
    \hline
    \end{tabular}
    \caption{\justifying
    \footnotesize Natural logarithm of the \acp{BF} and maximum likelihood for the logarithmically spaced spline model, relative to the 10-component model of the fixed-cosmology case. $\#$ indicates the number of spline components.}
    \label{tab:BFs}
\end{table}

\begin{figure*}
    \centering
    \includegraphics[width=\linewidth]{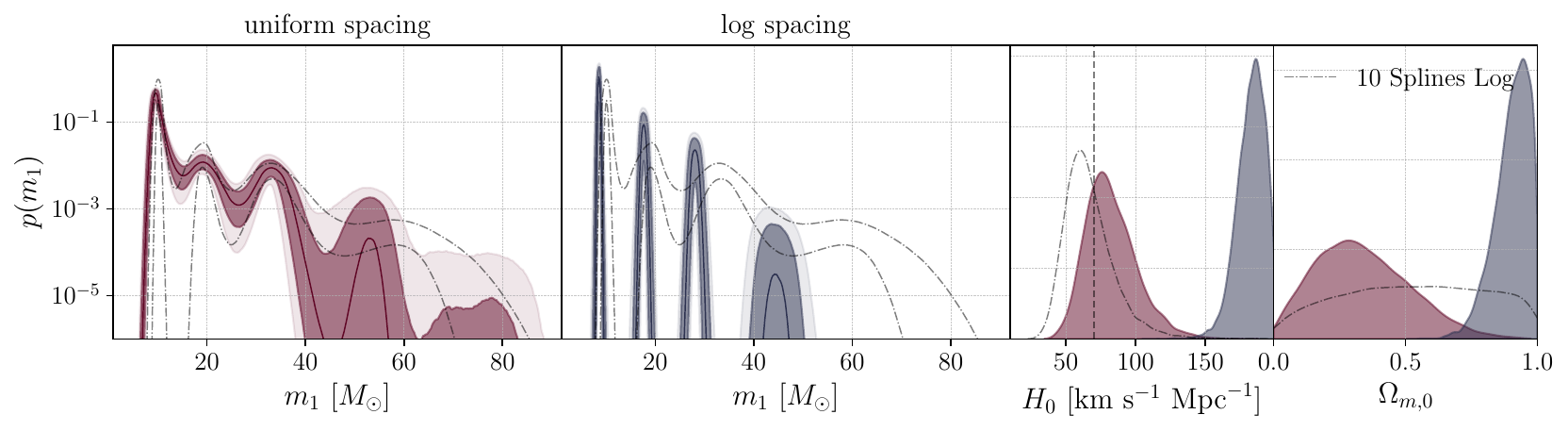}
    \caption{\justifying \footnotesize Reconstructed primary mass distributions using events from GWTC-4.0, modeled with a uniform (\textit{dark red}) and logarithmic (\textit{dark blue}) knot 20-spline, showing the median, $68\%$ and $90\%$ \acp{CI}. The dash-dotted curves correspond to the logarithmically spaced 10-spline analysis. The columns on the right show the $H_0$ and $\Omega_{m,0}$ posteriors, with the dashed vertical line indicating 70 for $H_0$.}
    \label{fig:gwtc-4_20-spline_cosmology}
\end{figure*}

\section{Subpopulation analysis}\label{app:sub-population_supplemental}

\begin{figure}
    \centering
    \includegraphics[width=0.38\textwidth]{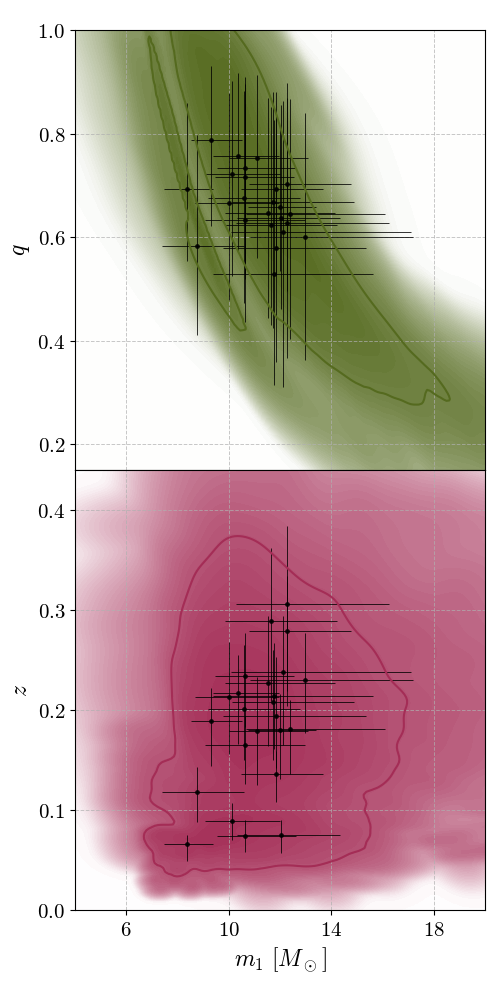}
    \vspace{-0.2cm}
    \caption{\justifying \footnotesize \ac{KDE} of the \ac{PE} posteriors for the 24 events around the $10M_{\odot}$ peak from GWTC-4.0. Crosses indicate the median values and the $68\%$ of the marginal distributions. Samples are converted to the source frame using the Planck cosmology~\cite{Planck:2018vyg}.}
    \label{fig:10M_posteriors_KDE}
\end{figure}

We list the 24 events used in the $10M_{\odot}$ analysis, whose posteriors are shown in Fig.~\ref{fig:10M_posteriors_KDE}:\\\vspace{-2mm} \newline
\GW{170608\_020116}, \GW{190707\_093326}, \GW{190725\_174728}, \GW{190728\_064510}, \GW{190924\_021846}, \GW{191103\_012549}, \GW{191105\_143521}, \GW{191129\_134029}, \GW{191204\_171526}, \GW{191216\_213338}, \GW{200202\_154313}, \GW{200316\_215756}, \GW{230627\_015337}, \GW{230630\_234532}, \GW{230729\_082317}, \GW{230731\_215307}, \GW{230904\_051013}, \GW{231018\_233037}, \GW{231020\_142947}, \GW{231104\_133418}, \GW{231113\_200417}, \GW{231223\_075055}, \GW{231223\_202619}, \GW{231224\_024321}.\newline\vspace{2mm}

\begin{figure*}[t]
    \centering

    \begin{subfigure}{0.49\textwidth}
        \centering
        \includegraphics[width=\linewidth]{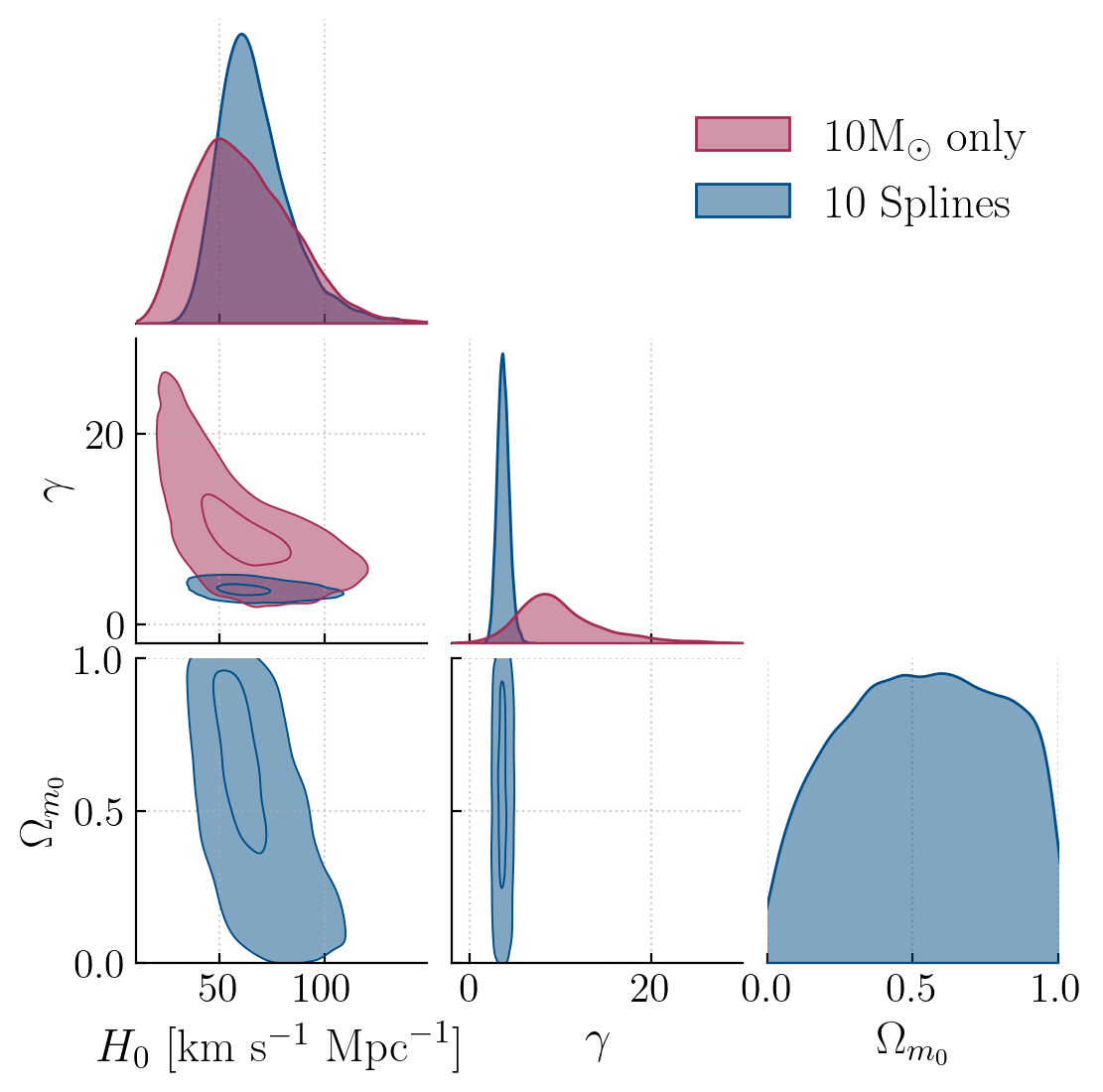}
    \end{subfigure}
    \hfill
    \begin{subfigure}{0.49\textwidth}
        \centering
        \includegraphics[width=\linewidth]{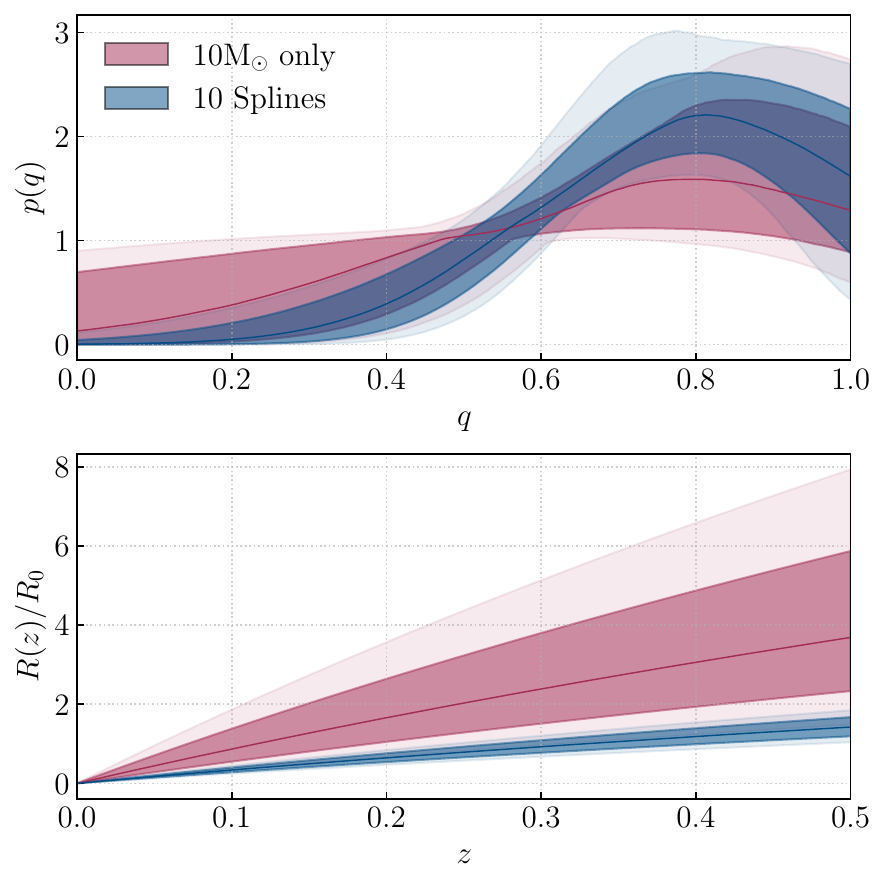}
    \end{subfigure}

    \caption{\justifying \footnotesize Results from the logarithmically spaced 10-spline analysis (blue) on the full \ac{BBH} population of GWTC-4.0 events, and from 4 uniform-spaced splines using 24 events around the $10M_{\odot}$ peak (pink). \textit{Left}: Corner plot showing posteriors for the rate evolution powerlaw index $\gamma$, and cosmological parameters $H_0$ and $\Omega_{m,0}$. \textit{Right}: Reconstructed mass ratio (top) and redshift evolution (bottom) distributions, showing the median, the $68\%$ and $90\%$ \acp{CI}.}
    \label{fig:GWTC-4_q-z}
\end{figure*}

\textit{Rate evolution of the subpopulation}. Fig.~\ref{fig:GWTC-4_q-z} compares the 10-spline logarithmic-spacing analysis on the full \ac{BBH} catalog with the $10M_{\odot}$ subpopulation analysis using 4 uniform-spaced splines. The left panel shows the posteriors for the cosmological parameters and the rate evolution powerlaw index, while the right panel presents the reconstructed mass ratio and redshift-dependent rate evolution. For the $10M_{\odot}$ subset, we adopt a simple powerlaw for the redshift dependence instead of the Madau–Dickinson model used for the full population and fix $\Omega_{m,0}=0.308$ to reflect the narrower, low-redshift range of these events (see Fig.~\ref{fig:10M_posteriors_KDE}).\newline

\begin{figure*}[t]
    \centering

    \begin{subfigure}{0.49\textwidth}
        \centering
        \includegraphics[width=\linewidth]{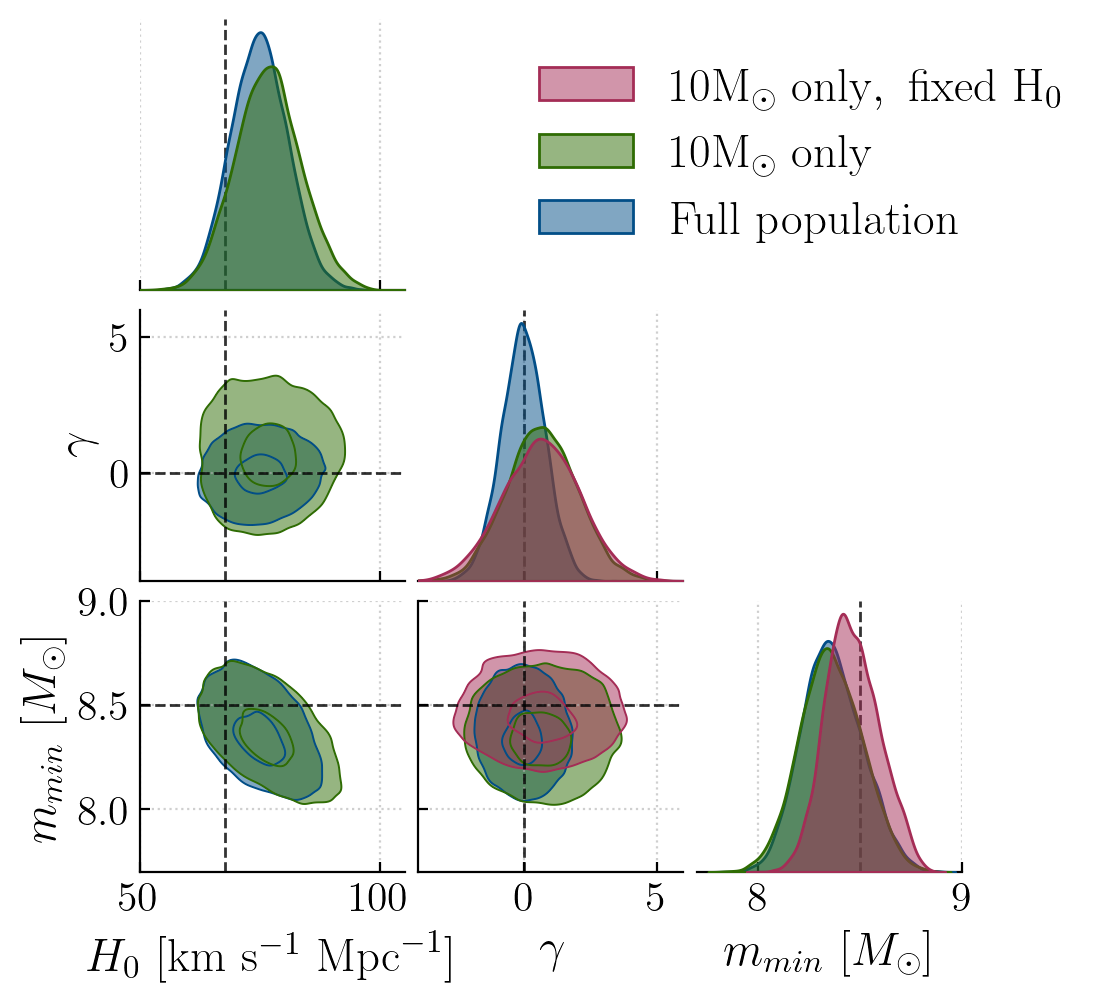}
    \end{subfigure}
    \hfill
    \begin{subfigure}{0.49\textwidth}
        \centering
        \includegraphics[width=\linewidth]{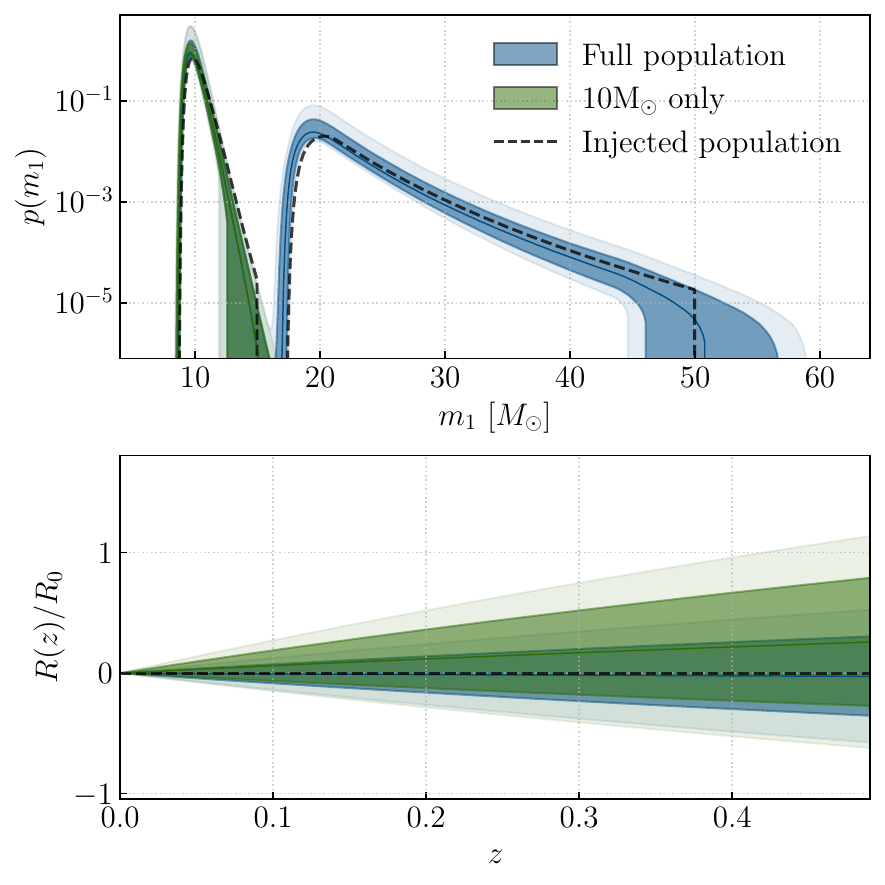}
    \end{subfigure}

    \caption{\justifying \footnotesize Simulated analysis of a synthetic population of 219 \acp{BBH} at O4 sensitivity. Events are drawn from a distribution composed of two tapered powerlaws (dashed black), and recovered using the same model (blue) and a low-mass subpopulation only (green). \textit{Left}: Corner plot showing posteriors for the Hubble constant $H_0$, rate evolution powerlaw index $\gamma$, and minimum mass $m_{min}$. Red contours correspond to the subpopulation analysis with cosmology fixed to the injected value. \textit{Right}: Reconstructed primary mass (top) and redshift evolution (bottom) distributions, showing the median, the $68\%$ and $90\%$ \acp{CI}.}
    \label{fig:GWTC-4_combined}
\end{figure*}

\textit{Simulated subpopulation analysis}. To validate our subpopulation approach, we simulate events from the population distribution in Fig.~\ref{fig:GWTC-4_combined}, composed of two separated tapered powerlaw components for the primary mass, broadly mimicking the GWTC-4.0 distribution and including the $15M_{\odot}$ gap separating low-mass events. We assume a flat powerlaw for the rate evolution and a Gaussian for the mass ratio centered at $q=0.8$ with $\sigma_{q}=0.1$, truncated at $q=1$. Using an \ac{LVK} detector network at O4 design sensitivity~\cite{KAGRA:2013rdx}, we obtain 219 detected events. Fig.~\ref{fig:GWTC-4_combined} reports population analyses using the true event values instead of \ac{PE} samples, reducing uncertainties to highlight potential systematic effects~\cite{Gennari:2025nho}. When recovered with the same population model, the injected parameters are unbiasedly retrieved. Using only the 142 events in the low-mass feature and applying a single tapered powerlaw model, the population parameters, including the rate evolution and $H_0$, are correctly inferred. The posterior on the rate evolution index is broader, as expected from the smaller sample ($\sim 60\%$ of the full catalog), and the $H_0$ constraints remain nearly identical. Finally, fixing the cosmology to the injected values yields consistent posteriors, confirming the robustness of the subpopulation analysis.

\section{Numerical stability and priors}\label{app:numerical_stability}

We use the effective number of \ac{PE} samples and injections to define the regions of parameter space where the hierarchical likelihood is numerically stable~\cite{Talbot:2023pex}. Specifically, we adopt thresholds of $10$ for the effective number of samples and $4 N_{\text{events}} = 600$ for the effective number of injections~\cite{Mastrogiovanni:2023zbw}. For the subpopulation analysis we adopt an even stricter cut on the log-likelihood variance, requiring it to be below one~\cite{Talbot:2023pex}. The priors on the coefficients of the log-\ac{PDF} B-spline model are chosen uniform in the range $[-70,150]$, which corresponds to the maximum interval allowing stable initialisation of the sampler for analyses with many splines, due to the numerical stability cuts. We generally observe strong positive correlations between coefficients, consistent with the near-degeneracy under global scaling described in App.~\ref{app:splines}. As a result, posteriors sometimes rail against the prior bounds depending on the coefficient and spline number. These issues could be mitigated through different modelling choices~\cite{Tagliazucchi:2026gxn}. We verified that the results are robust to changes in prior bounds and the number of live points. Finally, the analysis runtime grows rapidly with the number of splines, becoming prohibitive even on \acp{GPU}. New methods to accelerate these analyses will be needed in the near future.

\clearpage
\bibliography{references}

\end{document}